\documentclass[reprint, amsmath,amssymb,aps,floatix,
               superscriptaddress]{revtex4-2}

\usepackage{hyperref}
\usepackage{graphicx}
\usepackage{bm}
\usepackage[T1]{fontenc}
\usepackage[utf8]{inputenc}
\usepackage{amsmath,amssymb}
\usepackage{mathtools}
\usepackage{siunitx}
\usepackage{bm}
\usepackage{ae}
\usepackage{siunitx}
\usepackage{txfonts}
\usepackage{braket}
\usepackage{booktabs}

\newcommand{\eMax}{\ensuremath{e_{\text{max}}}}

\begin{document}

\preprint{APS/123-QED}

\title{Application of efficient generator-coordinate subspace-selection
algorithm to neutrinoless double-$\beta$ decay}

\author{A. M. Romero} \email{amromero@email.unc.edu}
 \affiliation{Department of Physics and Astronomy, University of North Carolina,
 Chapel Hill, North Carolina 27516-3255, USA}

\author{J. M. Yao} \email{yaojm8@mail.sysu.edu.cn}
\affiliation{School of Physics and Astronomy, Sun Yat-sen University, Zhuhai
519082, P.R. China}
\affiliation{Facility for Rare Isotope Beams, Michigan State University, East
Lansing, Michigan 48824-1321, USA}

\author{B. Bally}
\affiliation{Departamento de F\'isica Te\'orica y Universidad Aut\'onoma de
Madrid, E-28049 Madrid, Spain}

\author{T. R. Rodr\'iguez}
\affiliation{Departamento de F\'isica Te\'orica y Universidad Aut\'onoma de
Madrid, E-28049 Madrid, Spain}
\affiliation{Centro de Investigaci\'on Avanzada en F\'isica
Fundamental-CIAFF-UAM, E-28049 Madrid, Spain}

\author{J. Engel} \email{engelj@physics.unc.edu}
 \affiliation{Department of Physics and Astronomy, University of North Carolina,
 Chapel Hill, North Carolina 27516-3255, USA}

\date{\today} 

\begin{abstract}
The generator coordinate method begins with the variational construction of a
set of non-orthogonal mean-field states that span a subspace of the full
many-body Hilbert space.  These states are then often projected onto states with
good quantum numbers to restore symmetries, leading to a set with members that
can be similar to one another, and it is sometimes possible to reduce this set
without greatly affecting results.  Here we propose a greedy algorithm that we
call the energy-transition-orthogonality procedure (ENTROP) to select subsets of
important states.  As applied here, the approach selects on the basis of
diagonal energy, orthogonality, and contribution to the matrix element that
governs neutrinoless double-$\beta$ decay.  We present both shell-model and
preliminary \emph{ab initio} calculations of this matrix element for the decay
of $^{76}$Ge, with quadrupole deformation parameters and the isoscalar pairing
strength as generator coordinates.  ENTROP converges quickly, reducing
significantly the number of basis states needed for an accurate calculation. 
\end{abstract}

\maketitle

\section{Introduction}

The observation of neutrinoless double-$\beta$ ($0\nu\beta\beta$) decay, in
which two protons decay into two neutrons without neutrino emission, would show
that neutrinos are Majorana particles. The half life for such a decay depends on
the nuclear matrix element (NME) of the transition operator between the
ground-states of the initial and final nuclei.  The NME, which must be computed,
is model-dependent, with results differing from model to model by factors of up
to three. Reducing the discrepancy is difficult because each model has its own
uncontrolled approximations.  One way forward is to use \textit{ab initio}
methods to compute the NME from first principles. In particular, in-medium
similarity renormalization group (IMSRG) methods~\cite{imsrgI, imsrgII} with
chiral interactions are promising and have already been applied to nuclei such
as $^{48}$Ca \cite{yao2020ab,Belley2021} and $^{76}$Ge \cite{Belley2021} that
are of great interest to experimentalists.  The approach leads to effective
Hamiltonians and transition operators to be used together with traditional
many-body methods that cannot by themselves easily incorporate high-energy
correlations.  With the generator coordinate method (GCM)~\cite{hillwheeler} as
the traditional one, the approach has proved successful in describing the
spectra of low-lying states, and has been used to compute the NME for
$0\nu\beta\beta$ decay of $^{48}$Ca~\cite{yao2018generator,yao2020ab}.

The GCM, which has been applied most often within nuclear
energy-density-functional theory
\cite{rodriguez2010energy,Vaquero:2014dna,Song:2014,Yao:2015}, provides an
effective way to construct wave functions that include collective correlations.
Such correlations, in particular involving deformation (both axial and
triaxial~\cite{jiao}), and pairing (of both like-particle and
proton-neutron~\cite{hinohara,pairingII} type), are important for
$0\nu\beta\beta$ NMEs.   The GCM incorporates the effects of these degrees of
freedom by taking them as ``generator coordinates,'' with values on a mesh that
approximates the continuum.  Unfortunately, the method scales exponentially with
the number of coordinates. Including many mesh points leads to a large set of
non-orthogonal states and a significant computational burden.  Some of these
basis states, however, may closely resemble others or have little representation
in low-lying wave functions, and can therefore be omitted.  Here, we propose a
schemes that we call the energy-transition-orthogonality procedure (ENTROP) for
rejecting unimportant states.  As the name suggests, the approach is designed to
work for transition matrix elements and we apply it to $0\nu\beta\beta$ NMEs.

The particular case that we examine is the decay of $^{76}$Ge to $^{76}$Se. Both
nuclei exhibit triaxial deformation~\cite{toh2013, rodriguez2017role}.  If the
GCM includes two deformation coordinates and one that represents the effects of
isoscalar pairing, \cite{pairingII,jiao} in a large single-particle space, the
computing time required to restore all the broken symmetries in the resulting
set of states can be significant, making the three-coordinate case a good one
for testing/applying our algorithms. We do so within two kinds of calculations,
the first in a small shell-model valence space and an appropriate
semi-phenomenological interaction, and the second in seven major shells and an
\emph{ab initio} interaction resulting from the in-medium evolution of a chiral
Hamiltonian.  

The rest of this paper is organized as follows: Section~\ref{sec:method}
discusses the nature of the GCM basis and presents ENTROP, along with a
procedure based on the work of Ref.\ \cite{nakatsukasa}.  In
Section~\ref{sec:results} we present the results obtained after the application
of these methods in the computations just described. In
Section~\ref{sec:conclusions}, we offer conclusions.


\section{Methods}\label{sec:method}

$M_{0 \nu}$, the NME that we wish to calculate enters the rate of $0\nu\beta
\beta$ decay that is mediated by the exchange of light Majorana neutrinos as
follows:
\begin{equation}
\label{eq:bbrate}
\left[T_{1/2}^{0 \nu}\right]^{-1}=G_{0\nu}(Q,Z) 
\left| M_{0\nu} \right|^{2} \left|\sum_{k} m_{k} U_{ek}^{2}\right|^{2} \,, 
\end{equation}
where $Q$ is the energy difference between the initial and final atoms, $G_{0
\nu}$ is a phase space factor, the $m_{k}$ are the masses of the three light
neutrinos and the $U_{ek}$ are the elements of the neutrino mixing matrix that
connects the electron neutrino to the state with mass eigenvalue $m_{k}$.  One
traditionally separates $M_{0 \nu}$ into Gamow-Teller, Fermi, and tensor pieces,
\begin{equation}
\label{eq:m0v}
M_{0\nu} = M_{0\nu}^{GT} - \frac{g_{V}^2}{g_{A}^2}M_{0\nu}^{F} +
M_{0\nu}^{T} \,,
\end{equation}
where $g_{V}$ and $g_{A}$ are the nuclear vector and axial-vector weak coupling
constants (we use $g_{A} = 1.27$ here) and $M_{0\nu}^{GT}$, $M_{0\nu}^{F}$, and
$M_{0\nu}^{T}$ are defined, e.g., in Ref.\ \cite{em} (though $M_{0\nu}^{F}$
mistakenly contains an extra factor of $g_{V}^2/g_{A}^2$ there.)

The GCM combines constrained mean-field states into a fully-correlated nuclear
wavefunction~\cite{RS}, which we call a GCM state from now on. The starting
point is the set of mean-field states, for us Hartree-Fock-Bogoliubov (HFB)
quasiparticle vacua $\ket{\varphi(\bm{q})}$, that minimize the energy
$\bra{\varphi(\bm{q})} H \ket{\varphi(\bm{q})}$ under the constraint that a
vector of observables $\bm{\hat{Q}} = (\hat{Q}_1, \hat{Q}_2, \dots, \hat Q_N)$
takes the values $\bra{\varphi(\bm{q})}\bm{\hat{Q}}\ket{\varphi(\bm{q})} =
\bm{q}$.   The coordinates $\bm{q}$ that label the mean-field states are
frequently chosen to lie on an $N$-dimensional mesh that discretizes the space
of quasiparticle vacua.  The coordinate operators $\bm{\hat{Q}}$ are generally
those that are important for a good description of the nucleus.  In this paper,
we choose the quadrupole operators, $\hat{Q}_{20}$ and $\hat{Q}_{22}$, and the
isoscalar-pair creation operator $\hat{P}_0^+$ as generator coordinates.
Details appear in the next section.

A GCM state $\ket{\Psi^{JM}_{NZ}}$ has the form 
\begin{equation}
|\Psi^{JM}_{NZ}\rangle = \sum_{K,\bm{q}} f^{J}_{K,\bm{q}} \ket{NZJMK,\bm{q}} \,,
\end{equation}
where the states $\ket{NZJMK,\bm{q}}$ are projections of the $\ket{\varphi(\bm
q)}$: 
\begin{equation}
\ket{NZJMK,\bm{q}}=\hat{P}^J_{MK} \hat{P}^N \hat{P}^Z |\varphi(\bm{q})\rangle.
\end{equation}
Here $\hat{P}_{MK}^J$ is the operator that projects a state onto components with
well defined angular momentum $J$, $z$-projection $M$, and
intrinsic-$z$-projection $K$.  Since $K$ is not a good quantum number for a
triaxially-deformed nucleus, components with all values of $K$ contribute to a
GCM state (through ``$K$ mixing'').   The operators $\hat{P}^N$ and $\hat{P}^Z$
project states onto components with well-defined neutron number $N$ and proton
number $Z$. 

The projection operators produce basis states that are not orthonormal, and lead
to the Hill-Wheeler-Griffin (HWG) equation for $f_{K,q}^{J}$, 
\begin{equation}
\sum_{K',\bm{q}'} \big[ \mathcal{H}^J_{KK'}(\bm{q},\bm{q}') - E^J
\mathcal{N}^J_{KK'}(\bm{q},\bm{q}') \big] f^{J}_{K',\bm{q}'} = 0 \,,
\end{equation}
where the Hamiltonian and norm kernels $\mathcal{H}$ and $\mathcal{N}$ are given
by the expressions
\begin{eqnarray}
\mathcal{H}^J_{KK'}(\bm{q},\bm{q}') &=& \bra{NZJMK,\bm{q}} \hat{H}
\ket{NZJMK',\bm{q}'}\\ \label{eq:hamker}
\mathcal{N}^J_{KK'}(\bm{q},\bm{q}') &=& \braket{NZJMK,\bm{q} |NZJMK',\bm{q}'}
\,,
\label{eq:ovker}
\end{eqnarray}
and $E^J$ is the energy of the state with angular momentum $J$ that we are
interested in (we've suppressed the labels $N$ and $Z$ in places for
convenience).  We solve the HWG equation in the standard way \cite{RS}, by
diagonalizing the norm kernel to obtain a basis of ``natural states'' and then
diagonalizing the Hamiltonian $H$ in that basis.  The second diagonalization can
be numerically unstable, a problem we deal with by truncating the natural basis
to include only states with norm eigenvalues larger than a reasonable value.
That step eliminates the instability by removing states that are nearly linearly
dependent on others.

The computational time in this method lies mostly in the construction of the
kernels for the Hamiltonian and $0\nu\beta\beta$ transition operators.  That
process entails an integration of matrix elements of two-body operators over
Euler and gauge angles to project onto conserved quantities.  The norm kernels
require the same integration, though without an operator sandwiched between
states.  It is difficult to know ahead of time how dense to make the coordinate
mesh or how far to extend it, and so we would like to select a subset of points
on the mesh before computing all the kernels.  We can expect some basis states
to contribute little to the energy of the GCM ground state or to the
$0\nu\beta\beta$ NME between two GCM states, and others to be very similar to
one another (the result of too dense a mesh).  Our best prescription for subset
selection is based on three observations:
\begin{itemize}
\item States with lower expectation values for the Hamiltonian are in general
more important than those with higher expectation values. 
\item The largest contributions to NMEs often come from transitions between
basis states (in our case in two different nuclei) with the same values for the
collective coordinates $\bm{q}$ \cite{Menendez2011,rodriguez2010energy}.   
\item States that can nearly be represented as a linear combination of states in
the selected subset need not themselves be included in the subset.  They add
only numerical noise to the HWG equation that must be removed in its solution.
\end{itemize} 

\begin{figure}
\includegraphics[width=0.5\textwidth]{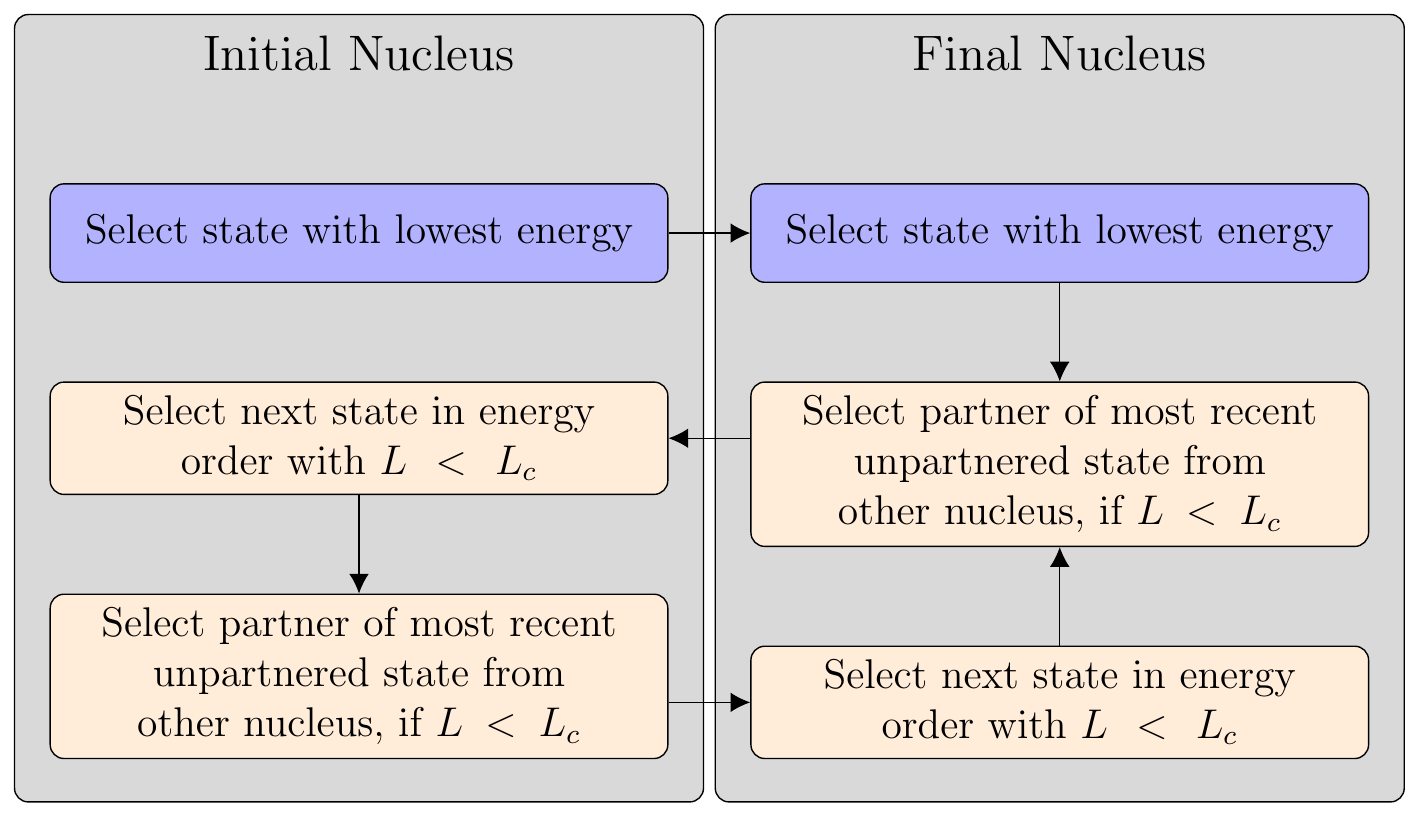}%
\caption{Schematic diagram of the state selection order produced by ENTROP.} 
\label{fig:algo_schematic}
\end{figure}

ENTROP incorporates these observations through the following procedure:  we
order the $\ket{\varphi(\bm{q})}$ in each nucleus by diagonal energies
$\braket{H}_{JKq}\equiv \mathcal{H}^J_{KK}(\bm{q},\bm{q})/
\mathcal{N}^J_{KK}(\bm{q},\bm{q})$ and select the one with the lowest value in,
e.g.,\ the initial nucleus.  We then move to the final nucleus, selecting first
the state with the lowest diagonal energy and then the state with the same
coordinates $\bm{q}$ as the first state from the initial nucleus (the
``partner'' of that state), provided that its projection onto the previously
included state has squared length $L$ less than some cutoff value $L_c$ (so that
it is nearly linearly independent).  Next we return to the initial
nucleus,selecting the state with the second-lowest diagonal energy and the
partner of the first the state in the final nucleus, again after checking
projections.  We continue in this way, including each state that we examine only
if its projection onto the space of previously-selected states has length less
than $L_c$, i.e.\ if
\begin{equation}
\label{eq:proj-defn}
L \equiv \frac{\braket{n + 1|P^{(n)}|n + 1}}{\braket{n + 1| n + 1} } < L_c.  
\end{equation}
Here $\ket{n + 1}$ is the state we are testing and $P^{(n)}$ is the projector
onto the $n$ states already selected (see the Appendix for details).  After
including each new state we diagonalize $H$ in the appropriate subset and look
for convergence of the eigenvalues and NME.  Fig.~\ref{fig:algo_schematic}
contains a flow chart representing the selection procedure.   The method saves
time because we compute the off-diagonal norm kernels only of the states we
examine and the off-diagonal Hamiltonian and $0\nu\beta\beta$ kernels only of
the states we eventually select.  

The procedure just outlined contains the parameter $L_{c}$, the value of which
we have yet to specify.  To determine it, we repeat the entire procedure for a
range of $L_{c}$ and within several pairs of small subspaces of the full space
(one space in the pair for the initial nucleus and one for the final).  We then
choose the smallest value of $L_{c}$ that ``works'' within each pair of
subspaces --- that is, a value that brings us so close to the energies and NMEs
obtained in each complete subspace pair that increasing $L_{c}$ further (and
thus including more basis states) has little effect.  We then assume that the
same will be true in any subspace pair, including one that contains all basis
states on the mesh in both nuclei.  This assumption cannot be rigorously
justified but is reasonable.
    
\begin{figure}
\includegraphics[width=0.5\textwidth]{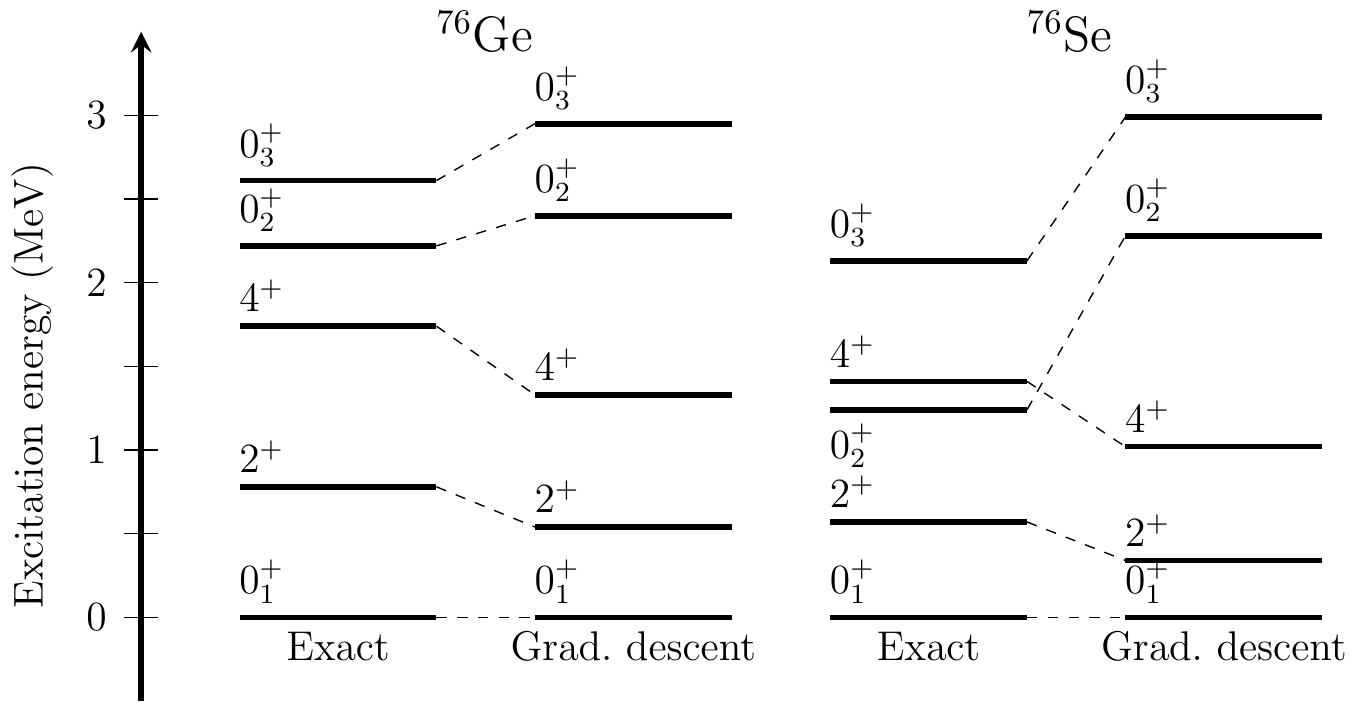}%
\caption{Low-lying energy spectrum of $^{76}$Ge and $^{76}$Se computed by
shell-model code BIGSTICK~\cite{bsI, bsII} (Exact) and from the procedure based
on that in the Ref.\ \cite{nakatsukasa} and described in the text (Grad.\
descent).  The figure doesn't show an overall upward shift in the Grad.\-descent
energies of about 1.5 MeV in $^{76}$Ge and 2 MeV in $^{76}$Se.}
\label{fig:spectrum}
\end{figure}

Our original intent was to implement something like the procedure discussed in
Ref.\ \cite{nakatsukasa}, which successfully reproduces the low-lying portions
of collective spectra within energy-density functional theory.  In that
approach, one starts from random mean-field states (Slater determinants in Ref.\
\cite{nakatsukasa} itself) obtained without constraints, descending towards
local minima in the energy surface via imaginary-time evolution and selecting
states along the way to subject to an orthogonality test like the one described
here.  We test a modification of that procedure, in which we use gradient
descent rather than imaginary-time evolution to approach energy minima in our
space of quasiparticle vacua, for the decay of $^{76}$Ge to $^{76}$Se with the
shell-model space and Hamiltonian described at the beginning of the next
section.  We use 50 randomly selected quasiparticle vacua as starting points,
and then select a random number of states along the corresponding paths of
descent once the energy has dropped below 10 MeV. In the most successful version
of this procedure, we then order the states by energy and fix a cutoff $L_{c}$
in the same way as with ENTROP.  But while we can roughly reproduce the exact
spectra of $^{76}$Ge and $^{76}$Se with about 30 states in each nucleus (from
about 17 distinct starting points in $^{76}$Ge and 18 in $^{76}$Se), as we show
in Fig.~\ref{fig:spectrum}, we are not able to obtain as accurate an NME as we
can with ENTROP (see Fig.~\ref{fig:nme_gcn}).  That result is not entirely
surprising because, unlike the GCM, the procedure of Ref.\ \cite{nakatsukasa} in
no way ensures that states in one nucleus are similar to those in the other. 

\section{Results}\label{sec:results}

\subsection{Shell-model test}

\begin{figure*}
\includegraphics[width=\textwidth]{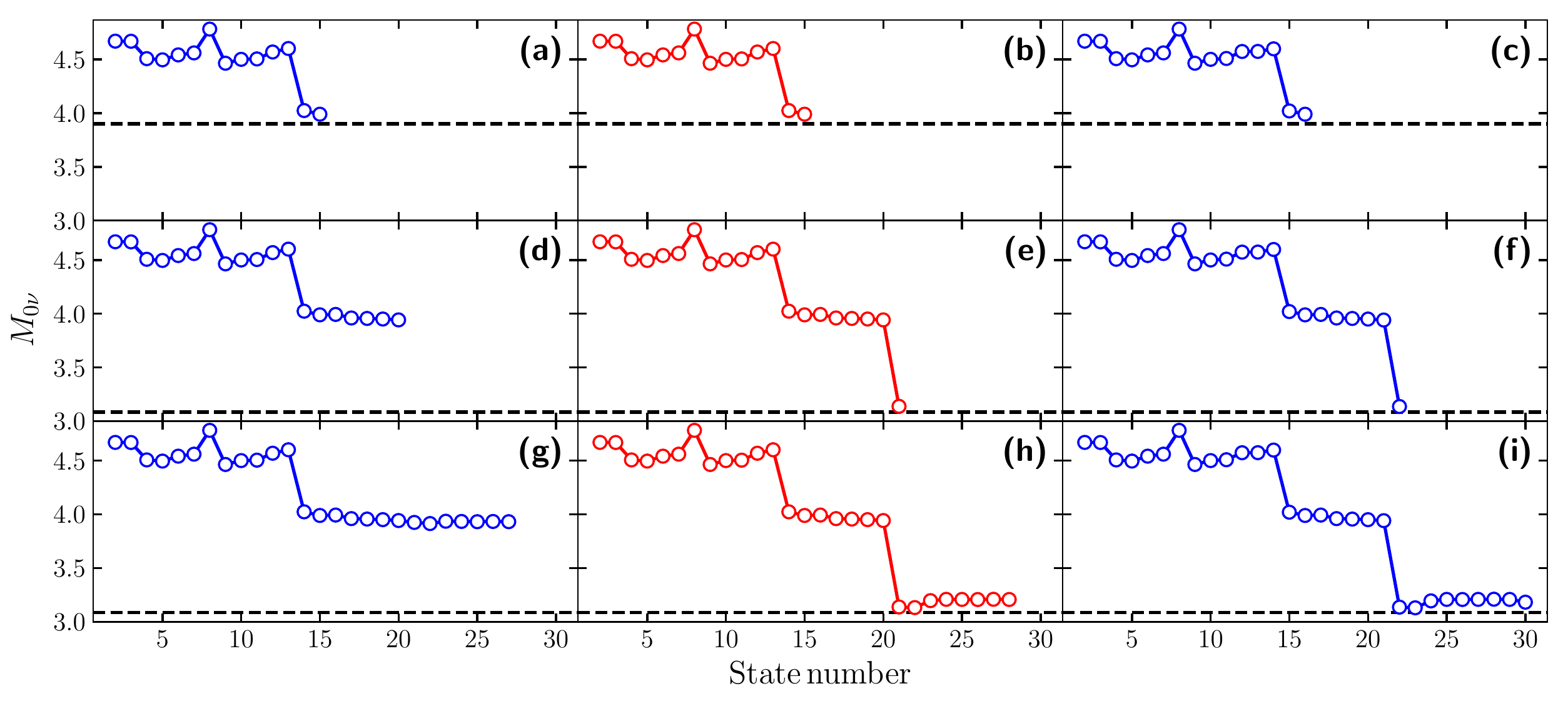}%
\caption{Valence-space NME for the decay of $^{76}$Ge, computed with the GCN2850
interaction in pairs of subspaces spanned together by the first 20, 40 and 60
states (top row, middle row, and bottom row) after applying ENTROP up to the
combined (from both-nuclei) state number indicated by the $x$ axis.  The states
are ordered as indicated in Fig.\ \ref{fig:algo_schematic}, with cutoff values
$L_{c}$ of 0.994 (left column), 0.995 (middle column), and 0.996 (right column).
The dashed line is the result produced by the full set of states in each
subspace pair.  The value $L_c = 0.995$, corresponding to the middle column (in
red) is the smallest that reproduces the full results in all three subspace
pairs.}
\label{fig:gcn_subset}
\end{figure*}

To test the accuracy of ENTROP, we examine the decay $^{76}$Ge $\longrightarrow$
$^{76}$Se in a model space built on the $0f_{5/2}$, $1p_{3/2}$, $1p_{1/2}$ and
$0g_{9/2}$ orbits, with the effective valence-space shell-model Hamiltonian
GCN2850~\cite{disassembling}.  The model space allows an exact solution with
modern shell model codes.  Ref.~\cite{jiao} carefully examined the performance
of the GCM for this problem, constructing a mesh of 184 quasiparticle vacua with
constraints on the coordinates representing axial deformation, triaxiality, and
the isoscalar pairing strength. The operators that correspond to these
coordinates are 
\begin{eqnarray}
\hat{Q}_{20} & = &\sum_i r_i^2 Y_i^{20} \nonumber \\
\hat{Q}_{22} & = &\sum_i r_i^2 Y_i^{22} \\
\hat{P}_0 & = &\frac{1}{2\sqrt{2}} \sum_{l,\alpha } \sqrt{2l+1}[a_{l,
\alpha}^\dag a_{l, \alpha}^\dag ]^{J=1,T=0}_{M=0,T_z=0} \ + \ h.c. \,, \nonumber 
\end{eqnarray} 
where $i$ labels nucleons in first quantization, the square brackets signify the
coupling of orbital angular momentum, spin, and isospin, and the operator $a_{l,
\alpha}^\dag$ creates a particle in the single-particle level with orbital
angular momentum $l$ and other quantum numbers specified by $\alpha$.  Here we
replicate the calculation of Ref.\ \cite{jiao} to test the results of
restricting ourselves to particular subsets of its states.  To construct the
basis states and solve the resulting eigenvalue problem, we use the FORTRAN
program TAURUS~\cite{taurus, taurusII}.  Fig.~\ref{fig:spectrum} shows the
low-lying spectra produced by the 184-state GCM and the method related to that
of Ref.\ \cite{nakatsukasa}.  As mentioned in the methodology section, in order
to choose the cutoff $L_{c}$, we evaluate the NME in subspace pairs with
increasing dimension, here those spanned by the first 20, 40, and 60 states
chosen in the order indicated in Fig.\ \ref{fig:algo_schematic}, with $L_{c}$
set to 1 to make sure no states are skipped.  We then find that $L_{c} = 0.995$
(so that trial states have to be almost completely expressible in terms of those
already selected to be rejected) is the smallest value that accurately allows us
to reproduce the NME in all three subspaces.

Fig.~\ref{fig:gcn_subset} shows how well the cutoff $L_{c} = 0.995$ works for
the NME in the subspaces just mentioned.  In all three cases it yields a number
very close to the complete ones, with little more than half the basis states in
the two larger subspace pairs.  This analysis leads us to expect that the states
we will discard with $L_{c} = 0.995$ in our complete calculation so nearly lie
in the spaces spanned by the states we will have already selected that they will
not alter the results.  

\begin{figure}[b]
\includegraphics[width=0.5\textwidth]{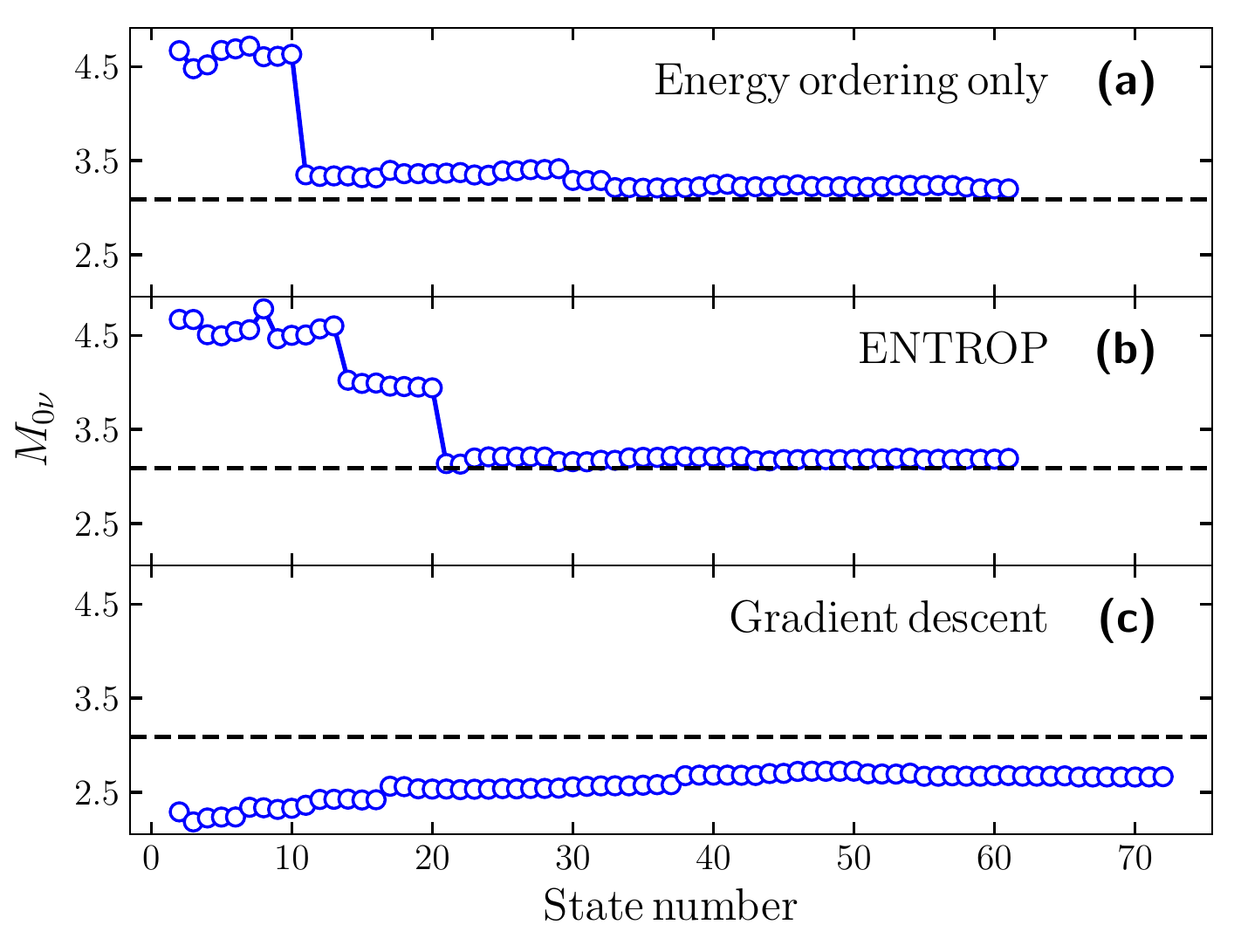}%
\caption{Valence-space NME for the decay of $^{76}$Ge produced by ENTROP without
including ``partner states'' (top, see text), by full ENTROP with $L_{c} =
0.995$ (middle), and by the procedure based on that in Ref.\ \cite{nakatsukasa}
(bottom), at the combined (both-nuclei) state number indicated by the $x$ axis.} 
\label{fig:nme_gcn}
\end{figure}

Our expectation turns out to be the case.  Fig.~\ref{fig:nme_gcn} shows the
results of our analysis in panel (b); after 20 states, the NME is very close to
the full GCM value.  Panel (a) in the same figure shows what happens when we do
not use the $0\nu\beta\beta$ operator to select states, that is, when we do not
include partner states.  Performance is generally worse, and even after 60
states the result is not as close to the full one as it is after 20 states in
panel (b).  Finally, panel (c) shows the result of the Ref.\
\cite{nakatsukasa}-like analysis discussed in the previous section.  As we noted
there, our NME does not approach the exact result within the set of states we
collect. 
   
\begin{figure}
\includegraphics[width=0.5\textwidth]{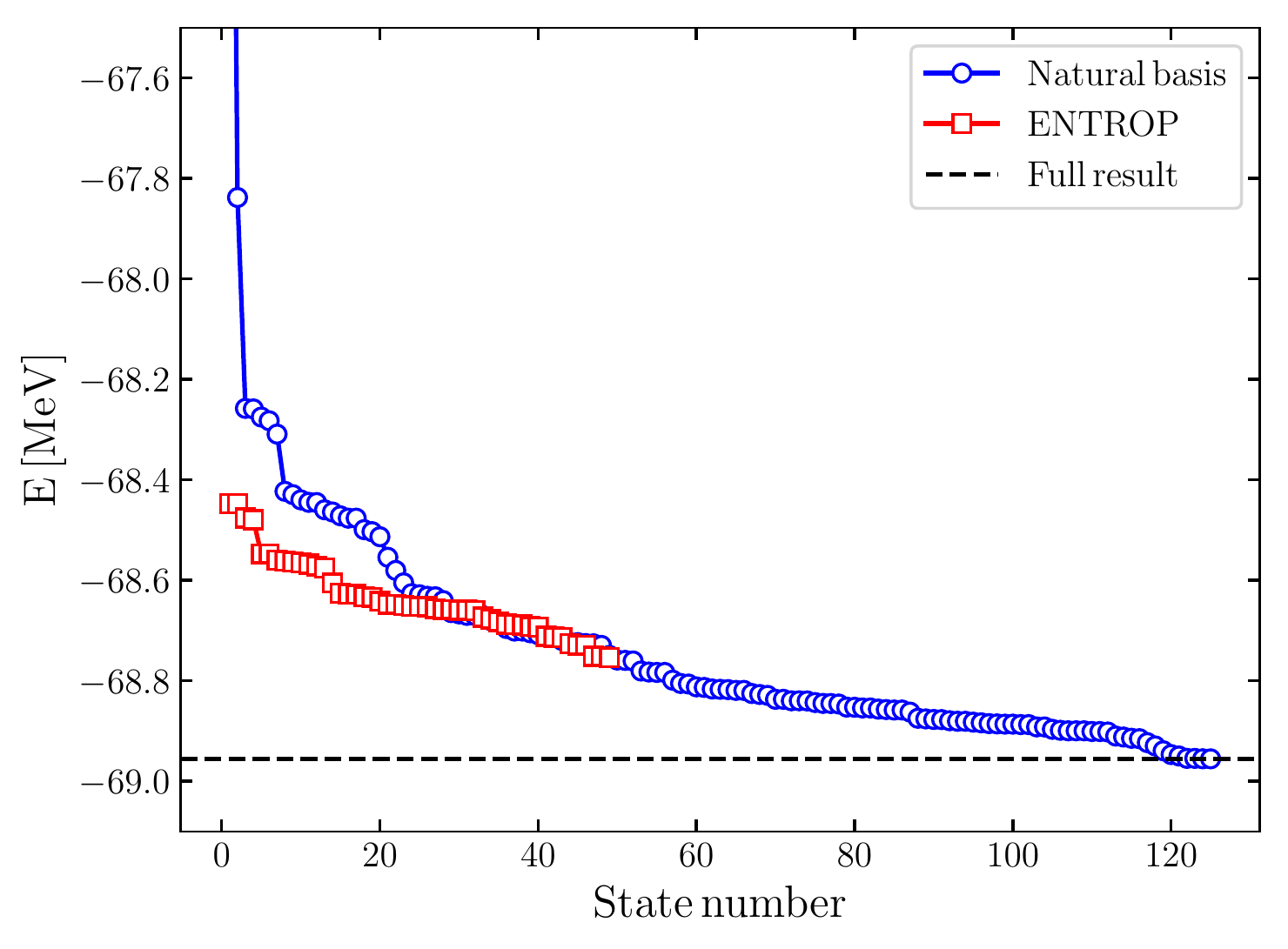}%
\caption{Valence-space $^{76}$Ge ground-state energy in the natural basis (blue)
and from ENTROP (red).  The dashed line is the full GCM result. Here, the state
number refers to a single nucleus only.}
\label{fig:energies_full_ge}
\end{figure}

\begin{figure}[b]
\includegraphics[width=0.5\textwidth]{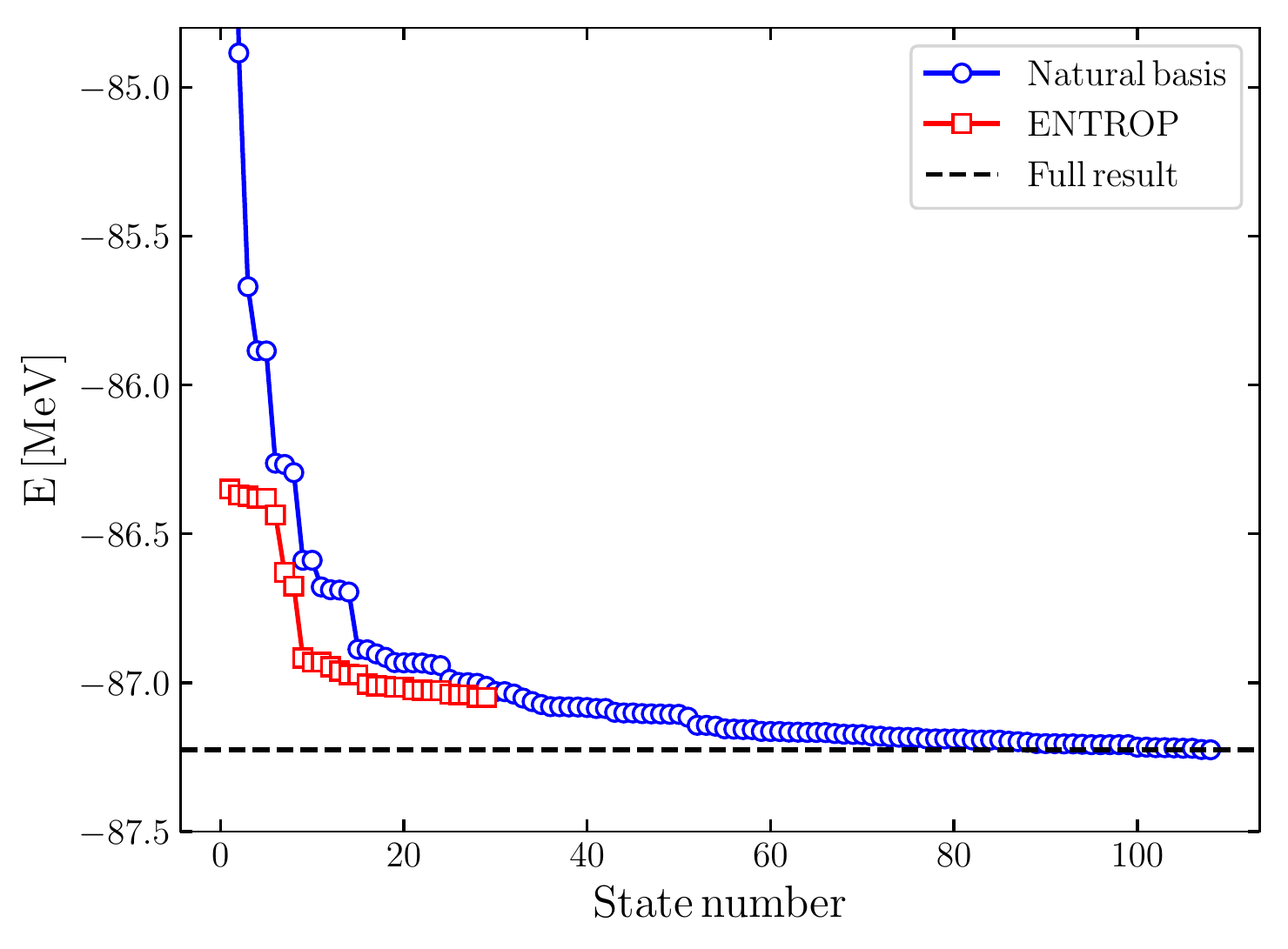}%
\caption{Same as Fig.\ \ref{fig:energies_full_ge} but for $^{76}$Se.}
\label{fig:energies_full_se}
\end{figure}

The convergence of the ground-state energies under ENTROP behaves a little
differently than that of the NME.  Figs.~\ref{fig:energies_full_ge} and
\ref{fig:energies_full_se} show the convergence towards the ground-state
energies of $^{76}$Ge and $^{76}$Se, respectively, within ENTROP and in the
full-GCM ``natural basis'', the one that for a given number of states picks out
the subspace that most closely spans the full set \cite{Srivastava_2000}.  Even
after the very first state --- the unconstrained HFB minimum, the ENTROP energy
is well within a percent of the correct one.  After that it converges more
gradually, eventually tracking the results of the natural-basis truncation.
Using a larger value of $L_{c}$ than 0.995 simply extends the ENTROP curve along
that corresponding to the natural basis.  We believe that this is the best that
one can do without an explicit (and time consuming) consideration of
off-diagonal contributions to the energy.  Fortunately, however, the long tail
of rejected states makes almost no difference in the NME; if we extend the
curves in the top two panels of Fig.\ \ref{fig:nme_gcn} the NME never moves
significantly from the full GCM value.

\subsection{\emph{Ab initio} calculation}

We turn now to the \emph{ab initio} computation of the same decay NME. Using a
chiral NN + 3N interaction \cite{Entem:2003,Hebeler:2011} employed in recent
studies of light nuclei \cite{Yao2021:benchmark} and $^{48}$Ca \cite{yao2020ab},
and evolving it and the decay operator according to the equations of the IMSRG
\cite{imsrgI} with a reference ensemble comprising prolate, spherical, and
oblate HFB minima in both $^{76}$Ge and $^{76}$Se and with $e_{\rm{max}}=6$
(i.e.\ in 7 shells), we repeat the steps just described.  Results in a larger
space will be published soon.  Unlike in our shell-model computation --- and
this would be the case in any realistic application --- we do not have
``complete'' results with which to test our approximations.  Our mesh in the
space of deformation parameters $\beta, \gamma$, and $\varphi$~\cite{jiao}
(related to the axial deformation, triaxiality and isoscalar pairing strength
used in the shell-model calculation) contains 145 points (or 290 if we count the
points in both nuclei), and a complete solution to the HWG equation in the
resulting space is more than we can currently handle.  We thus once again apply
ENTROP, this time without comparing to an exact result.

\begin{figure}
\includegraphics[width=0.5\textwidth]{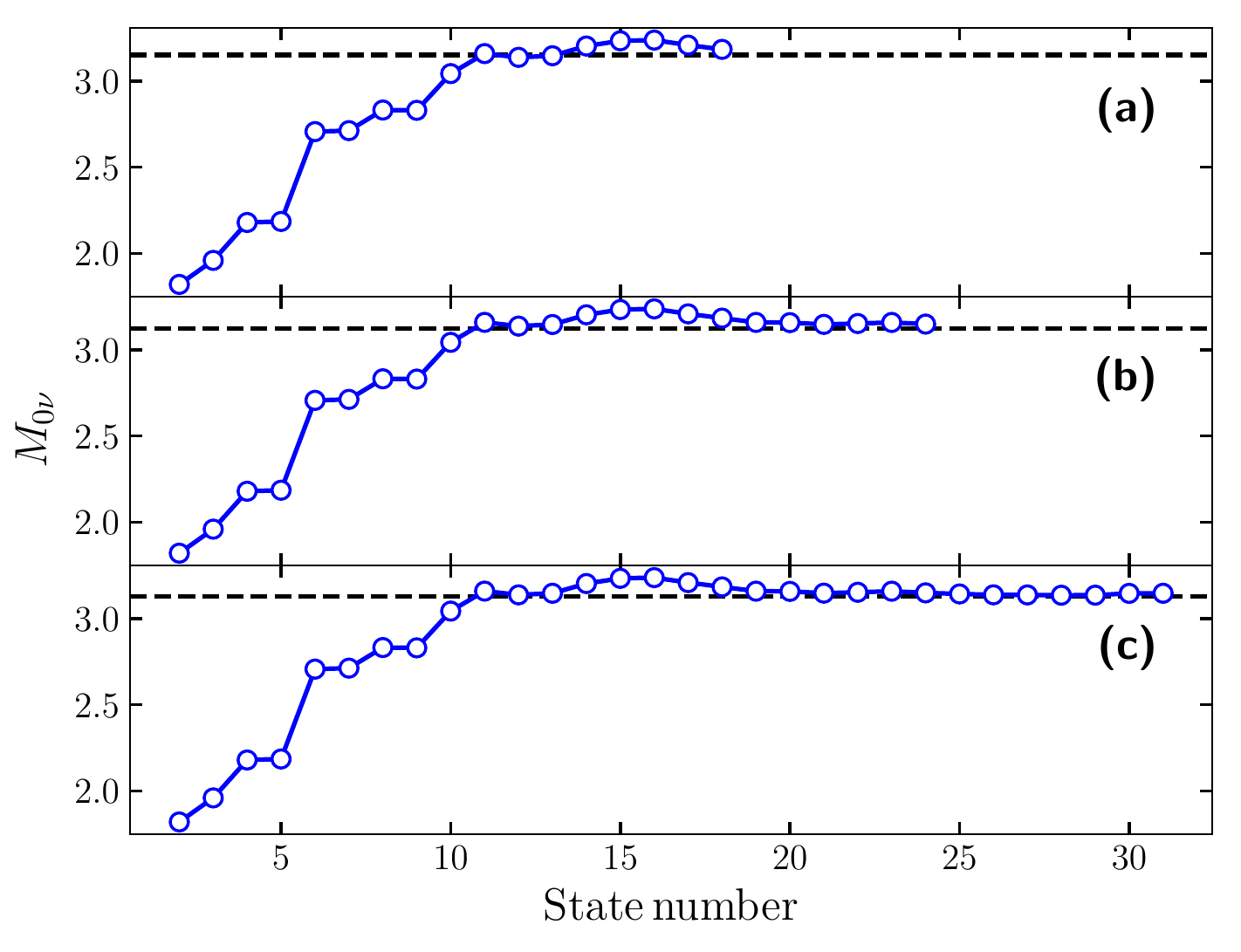}%
\caption{\textit{Ab initio} NME for the decay of $^{76}$Ge, with the total
number of states in the subset pairs equal to 20 (top), 30 (middle), and 40
(bottom), after applying ENTROP up to the combined (both-nuclei) state number
indicated by the $x$ axis.}
\label{fig:emax_grid_plot}
\end{figure}

\begin{figure}[b]
\includegraphics[width=0.5\textwidth]{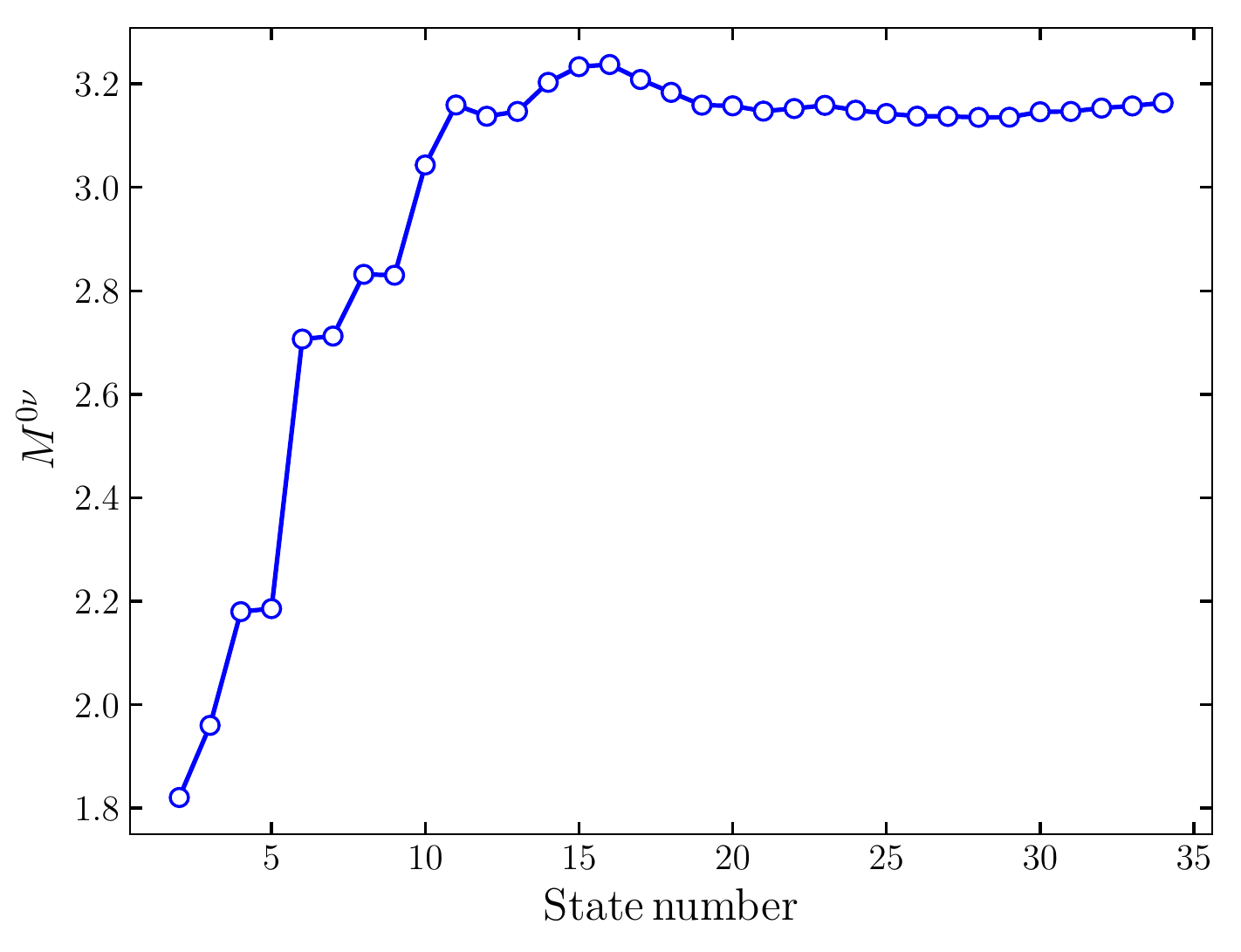}%
\caption{ENTROP NME for the decay of $^{76}$Ge, with $L_{c} = 0.902$.} 
\label{fig:NME_emax6}
\end{figure}

\begin{figure}
\includegraphics[width=0.5\textwidth]{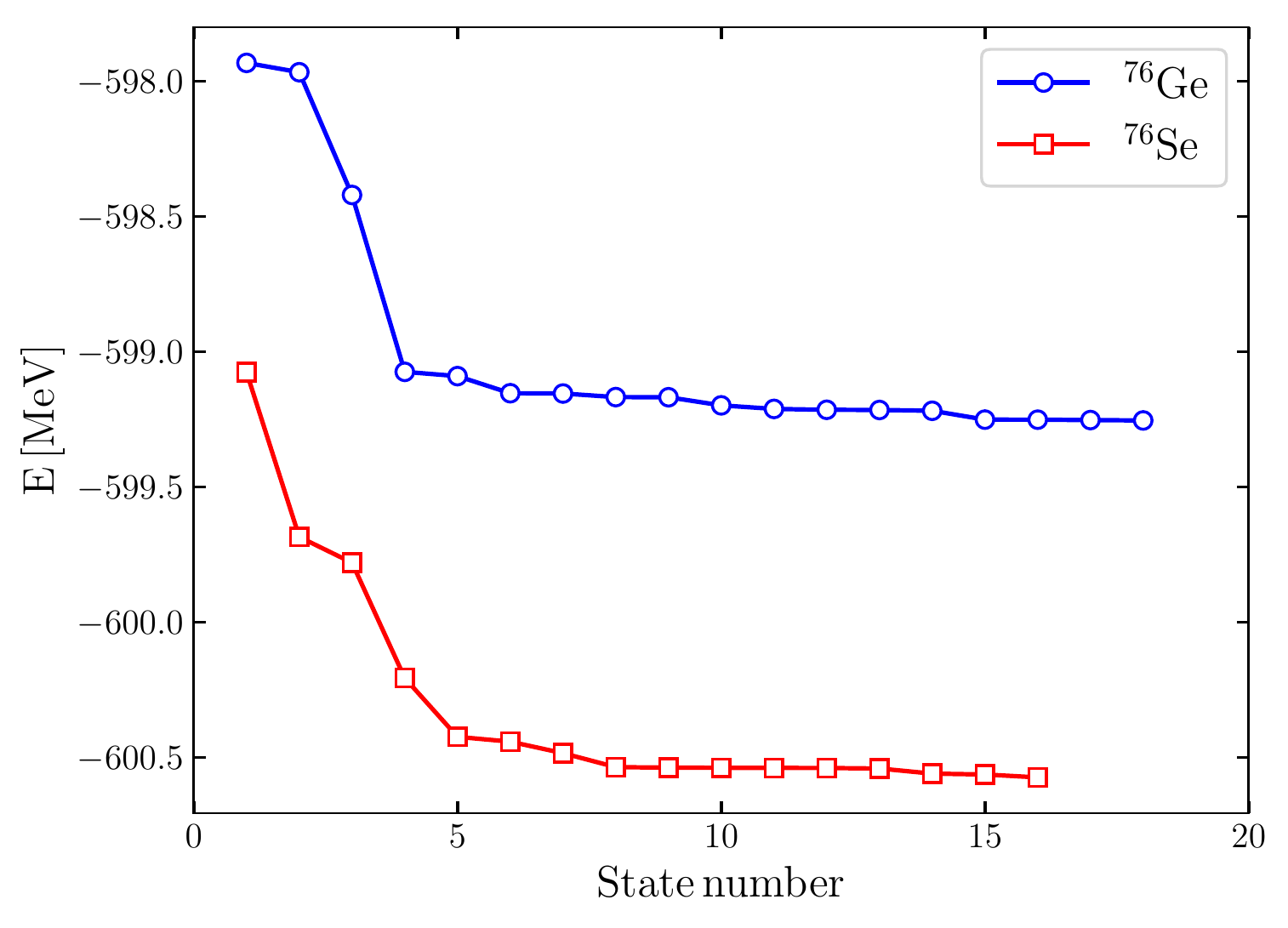}%
\caption{\textit{Ab initio} ENTROP ground-state energies for $^{76}$Ge (blue)
and $^{76}$Se (red), with $L_{c} = 0.902$.  The state number refers to a single
nucleus only.} 
\label{fig:energies_emax6}
\end{figure}

Fig.~\ref{fig:emax_grid_plot} shows that within subspace pairs consisting of 20,
30, and 40 total states, a cutoff value $L_{c} = 0.902$ is sufficient to obtain
the correct NME for each pair. It is the smallest value of the cutoff that does
so.  We therefore adopt this cutoff and generate another sequence of states,
leading to the results in Figs.~\ref{fig:NME_emax6} and
\ref{fig:energies_emax6}.  Though the energies in Fig.\ \ref{fig:energies_emax6}
are still falling slowly after 18 and 16 states in $^{76}$Ge and $^{76}$Se, the
NME in Fig.\ \ref{fig:NME_emax6} has more or less converged long before, by
about 20 states from the two nuclei combined.  Of course, we cannot be sure that
the long plateau continues indefinitely, but the longer it extends, the more
confidence we have.

The three parts of our NME are 
\begin{equation}
\label{eq:result}
\begin{aligned}
 M_{0\nu}^{GT}  &= 2.68 \\
 -\frac{g_{V}^2}{g_{A}^2} M_{0\nu}^{F} \, &= 0.65\\
 M_{0\nu}^{T} \, &= -0.16\,. 
\end{aligned}
\end{equation}
A recent valence-space IMSRG calculation obtained $M_{0\nu}^{GT} = 2.76$,
$g_{V}^2 / g_{A}^2 M_{0\nu}^{F} = 0.54$, and $M_{0\nu}^{T} = -0.49$ with the
same chiral interaction and the same value of $\eMax$ \cite{Belley2021}.  The
main difference between the two sets of results is in the tensor matrix element.
In the valence-space calculation, this component reduces the total NME by 17\%,
while in ours it reduces it by only 5\%, a number that is similar to what has
been obtained in more phenomenological computations.  We will publish a more
complete calculation of these matrix elements with a larger value for $\eMax$
elsewhere.

\section{Conclusions}\label{sec:conclusions}

We have presented a greedy algorithm called ENTROP to select the most important
mean-field states for GCM calculations of ground-state energies and
$0\nu\beta\beta$ NMEs.  The algorithm starts with one HFB quasiparticle vacuum
per point in a large mesh within a space of collective coordinates, and reduces
the number of projected HFB states that need to be worked with.  The steps in
the procedure, briefly, are as follows:
\begin{itemize}
\item Sort the projected states by their diagonal energies.
\item Consider the first $N$ states in each nucleus for several values of $N$.
\item Find the smallest value of $L_c$ that, when the selection scheme in Fig.\
\ref{fig:algo_schematic} is applied, leads to subsets of the first $N$ states
(for all the values of $N$) that succeed in reproducing the corresponding NME.
\item Use that value of $L_c$ to create a subspace pair in the full GCM spaces,
solve the corresponding HWG equations, and compute the NME.
\end{itemize}

The scheme reduces computational effort because we need to compute norm kernels
only for the projected states states that we test, and Hamiltonian and $\beta
\beta$ kernels only for those that are actually selected.  We successfully
tested our method in a computation of the NME for the decay of $^{76}$Ge within
a valence shell-model space with a phenomenological interaction; it reduced
computation time there by more than a factor of 100.  We also applied the method
to an \emph{ab initio} computation of the same NME with an IMSRG-evolved chiral
interaction, where a full calculation is too time consuming to carry out.  In
both our examples, ENTROP appears to lead to a suitable basis with many fewer
states than in typical GCM calculations, opening up the possibility of adding
new generator coordinates to the usual set.  

As we just noted, ENTROP requires norm kernels for the set of states that are
tested, and although those take less time to compute than do Hamiltonian or
$\beta \beta$ kernels, they are still not always cheap.  We have found the use
of approximate norm kernels, e.g.\ from unprojected basis states to be
promising, and are also exploring machine-learning techniques to reduce the
number of norm kernels that must be calculated. 

\begin{acknowledgments}
We thank H. Hergert, C.F. Jiao, and R. Wirth for fruitful discussions, and A.
Belley for sending us results of VS-IMSRG calculations.  This work is supported
in part by the U.S.  Department of Energy, Office of Science, Office of Nuclear
Physics under Awards No. DE-SC0017887, No.  DE-FG02-97ER41019, No.  DE-SC0015376
(the DBD Topical Theory Collaboration) and No. DE-SC0018083 (NUCLEI SciDAC-4
Collaboration).  It is also supported by the Spanish Ministerio de Ciencia e
Innovación under contract PGC2018-094583-B-I00 and the European Union’s Horizon
2020 research and innovation programme under the Marie Skłodowska-Curie grant
agreement No. 839847.  Computing resources were provided by the Institute for
Cyber-Enabled Research at Michigan State University, the Research Computing
group at the University of North Carolina, and the U.S. National Energy Research
Scientific Computing Center (NERSC), a DOE Office of Science User Facility
supported by the Office of Science of the U.S.  Department of Energy under
Contract No. DE-AC02-05CH11231.
\end{acknowledgments}

\appendix*

\section{Squared length of projection onto a subspace}\label{sec:app}

To compute $L$ for a given state and a subspace of previously selected states we
proceed as follows. Let the subspace be spanned by the un-normalized and
non-orthogonal vectors $\ket{1}, \ket{2}, \hdots, \ket{n}$.  A candidate state
$|n+1\rangle$ will not be included in this set if it is nearly a linear
superposition of those states. Calling the projector onto the subspace
$P^{(n)}$, we have
\begin{equation}
\label{eq:Pn1}
P^{(n)} |n+1\rangle = \sum_{i=1}^n \alpha_i^{(n)} |i\rangle,
\end{equation}
for some coefficients $\alpha_i^{(n)}$, which are determined by requiring that
$|n+1\rangle - P^{(n)}|n+1\rangle$ is orthogonal to $|k\rangle$ for all $k \leq
n$,
\begin{equation}
\label{eq:Pn1me}
\braket{k|n+1} - \braket{k|P^{(n)}|n + 1} = 0 \longrightarrow \sum_{i=1}^n 
\braket{k|i} \alpha_i^{(n)} = \braket{k|n + 1} \,. 
\end{equation} 
In matrix form, Eq.\ \eqref{eq:Pn1me} is 
\begin{equation}
\bm{S}^{(n)}\bm{\alpha}^{(n)}=\bm{\beta}^{(n)},
\end{equation}
with $S_{ij}^{(n)}= \braket{i|j}$ and $\beta_i^{(n)}= \braket{i|n + 1}$. The
solution is 
\begin{equation}
\bm{\alpha}^{(n)}=(\bm{S^{(n)}})^{-1} \bm{\beta^{(n)}} \,. 
\end{equation}
The squared length $L$ of the projection of the normalized candidate state onto
the space spanned by the already selected states is then 
\begin{eqnarray}
L &=& \frac{\braket{n + 1|P^{(n)}|n + 1}}{\braket{n + 1|n + 1}} =
\frac{\sum_{i=1}^n \braket{n + 1| i} \alpha_i^{(n)}}{\braket{n + 1|n + 1}}
\nonumber\\
&=& \frac{\bm{\beta}^{(n)\dag}(\bm{S}^{(n)})^{-1}\bm{\beta}^{(n)}} { \braket{n +
1|n + 1}} \,. 
\end{eqnarray}

\end{document}